\PassOptionsToPackage{dvipsnames}{xcolor}
\documentclass[sigconf,screen,authorversion]{acmart}

\usepackage[all]{nowidow}
\usepackage{xcolor}
\usepackage{caption}
\usepackage{subcaption}
\usepackage{hyperref}
\usepackage{acronym}
\usepackage{pifont}
\usepackage{fnpct}
\usepackage{mathtools}
\usepackage{bm}

\usepackage{booktabs}
\usepackage{makecell} 
\usepackage[flushleft]{threeparttable}
\usepackage{multirow}
\usepackage{tabularx}
\usepackage{tikz}
\usetikzlibrary{patterns.meta}

\newcommand{\hatchbox}{
\begin{tikzpicture}
    \draw (0,0) rectangle (0.3,0.3);
    \draw (0,0) -- (0.3,0.3);
    \draw (0.1,0) -- (0.3,0.2);
    \draw (0.2,0) -- (0.3,0.1);
    \draw (0,0.1) -- (0.2,0.3);
    \draw (0,0.2) -- (0.1,0.3);
\end{tikzpicture}
}

\newcommand{\linesbox}{
\begin{tikzpicture}
    \draw (0,0) rectangle (0.3,0.3);
    \draw (0,0) -- (0.3,0.3);
\end{tikzpicture}
}

\newcommand{\emptybox}{
\begin{tikzpicture}
    \draw (0,0) rectangle (0.3,0.3);
\end{tikzpicture}
}

\newcommand{\ldotsbox}{
\begin{tikzpicture}
    \draw (0,0) rectangle (0.3,0.3);
    \node at (0.05,0.05)[circle,fill,inner sep=0.25pt]{};
    \node at (0.05,0.117)[circle,fill,inner sep=0.25pt]{};
    \node at (0.05,0.183)[circle,fill,inner sep=0.25pt]{};
    \node at (0.05,0.25)[circle,fill,inner sep=0.25pt]{};

    \node at (0.117,0.05)[circle,fill,inner sep=0.25pt]{};
    \node at (0.117,0.117)[circle,fill,inner sep=0.25pt]{};
    \node at (0.117,0.183)[circle,fill,inner sep=0.25pt]{};
    \node at (0.117,0.25)[circle,fill,inner sep=0.25pt]{};

    \node at (0.183,0.05)[circle,fill,inner sep=0.25pt]{};
    \node at (0.183,0.117)[circle,fill,inner sep=0.25pt]{};
    \node at (0.183,0.183)[circle,fill,inner sep=0.25pt]{};
    \node at (0.183,0.25)[circle,fill,inner sep=0.25pt]{};

    \node at (0.25,0.05)[circle,fill,inner sep=0.25pt]{};
    \node at (0.25,0.117)[circle,fill,inner sep=0.25pt]{};
    \node at (0.25,0.183)[circle,fill,inner sep=0.25pt]{};
    \node at (0.25,0.25)[circle,fill,inner sep=0.25pt]{};
\end{tikzpicture}
}

\newcommand{\sdotsbox}{
\begin{tikzpicture}
    \draw (0,0) rectangle (0.3,0.3);
    \node at (0.15,0.15)[circle,fill,inner sep=0.25pt]{};
    \node at (0.05,0.05)[circle,fill,inner sep=0.25pt]{};
    \node at (0.05,0.25)[circle,fill,inner sep=0.25pt]{};
    \node at (0.25,0.05)[circle,fill,inner sep=0.25pt]{};
    \node at (0.25,0.25)[circle,fill,inner sep=0.25pt]{};
\end{tikzpicture}
}

\newcolumntype{Y}{>{\centering\arraybackslash}X}

\usepackage[capitalise,noabbrev]{cleveref}
\crefformat{section}{\S#2#1#3} 
\crefformat{subsection}{\S#2#1#3}
\crefformat{subsubsection}{\S#2#1#3}

\begin{document}

\author{Nils Japke}
\affiliation{%
    \institution{TU Berlin \& ECDF}
    \city{Berlin}
    \country{Germany}}
\email{nj@mcc.tu-berlin.de}

\author{Christoph Witzko}
\affiliation{%
    \institution{TU Berlin \& ECDF}
    \city{Berlin}
    \country{Germany}}
\email{chwi@mcc.tu-berlin.de}

\author{Martin Grambow}
\affiliation{%
    \institution{TU Berlin \& ECDF}
    \city{Berlin}
    \country{Germany}}
\email{mg@mcc.tu-berlin.de}

\author{David Bermbach}
\affiliation{%
    \institution{TU Berlin \& ECDF}
    \city{Berlin}
    \country{Germany}}
\email{db@mcc.tu-berlin.de}

\title[Studying Performance Issues Using Micro- and Application Benchmarks]{The Early Microbenchmark Catches the Bug -- Studying Performance Issues Using Micro- and Application Benchmarks}


\copyrightyear{2023}
\acmYear{2023}
\setcopyright{acmlicensed}\acmConference[UCC '23]{2023 IEEE/ACM 16th International Conference on Utility and Cloud Computing}{December 4--7, 2023}{Taormina (Messina), Italy}
\acmBooktitle{2023 IEEE/ACM 16th International Conference on Utility and Cloud Computing (UCC '23), December 4--7, 2023, Taormina (Messina), Italy}
\acmPrice{15.00}
\acmDOI{10.1145/3603166.3632128}
\acmISBN{979-8-4007-0234-1/23/12}

\begin{abstract}
An application's performance regressions can be detected by both application or microbenchmarks.
While application benchmarks stress the \textit{system under test} by sending synthetic but realistic requests which, e.g., simulate real user traffic, microbenchmarks evaluate the performance on a subroutine level by calling the \textit{function under test} repeatedly.

In this paper, we use a testbed microservice application which includes three performance issues to study the detection capabilities of both approaches.
In extensive benchmarking experiments, we increase the severity of each performance issue stepwise, run both an application benchmark and the microbenchmark suite, and check at which point each benchmark detects the performance issue.
Our results show that microbenchmarks detect all three issues earlier, some even at the lowest severity level.
Application benchmarks, however, raised false positive alarms, wrongly detected performance improvements, and detected the performance issues later.
\end{abstract}

\maketitle

\section{Introduction}
\label{sec:introduction}

In large-scale cloud systems, software performance issues can easily remain undetected and slip into production systems without proper care.
To avoid violating Service Level Agreements (SLAs), developers typically run benchmarks to assess performance of their systems or system under test (SUT)~\cite{grambow_continuous_2019,vanhoorn2012kieker}.
In general, there are two basic benchmarking variations, namely \emph{application benchmarks} and \emph{microbenchmarks}~\cite{book_bermbach2017_cloud_service_benchmarking}.
While application benchmarks set up the entire software system that should be evaluated and take the \emph{outer} user perspective when sending artificial requests to stress the SUTs, microbenchmarks do a fine grained evaluation on function\footnote{Generally, microbenchmarks are used with any kind of subroutine. In this paper, we will refer to them as functions.} level~\cite{grambow2021using}.
By repeatedly calling single functions and measuring the execution duration, a suite of microbenchmarks covering the complete source code can evaluate the same SUT from an \emph{inner} perspective.

Although both techniques are used in practice, a detailed comparison of which benchmark type can detect which types of performance issues faster and or more accurately is still missing.
One of the reasons for this is the lack of access to real world applications and their benchmarks for researchers.
In this paper, we close this gap and design a cloud-based testbed for studying performance issues.
For the testbed, we implemented a flight booking application as a microservice with three performance issues with configurable severity (additional ones can easily be added), an application benchmark, and a comprehensive suite of microbenchmarks, all with complete experiment automation.
We publish all components as open source\footnote{\url{https://github.com/njapke/flight-booking-service}}\footnote{\url{https://github.com/njapke/cloud-benchmark-conductor}}.
Then, we use the testbed to assess and compare the detection capabilities of application and microbenchmarks to answer the following research question:
Which benchmarking approach can detect which performance issues at which severity level and with which confidence level?

Overall, we make the following contributions:
\begin{itemize}
    \item We design and implement a flight booking application with three configurable performance issues as benchmarking testbed.
    \item We design and implement a realistic application benchmark, simulating end user requests.
    \item We design and implement a comprehensive microbenchmark suite covering the source code of the testbed application.
    \item We run extensive experiments using both benchmark types following benchmarking best practices and publish the resulting dataset.
		\item We comprehensively analyze and compare the detecting capabilities of both benchmark types.
\end{itemize}

Our findings show that microbenchmarks are much more effective at detecting performance issues than application benchmarks.
For all three performance issues, at least one microbenchmark consistently detects the performance issue at lower severity levels than the application benchmark.
Furthermore, the application benchmark tended to detect false positives, i.e., performance changes that do not exist, much more than the microbenchmarks.
In particular, even though we only inject performance regressions, the application benchmark sometimes measured performance improvements, i.e., a speedup of the version with the performance issue.
While microbenchmarks detect the performance issues much more effectively, they cost more to execute.
In our studied setup, we needed three Cloud VMs for the microbenchmark suite with an experiment duration of approx.\ one hour, while the application benchmark needed just two Cloud VMs for approx.\ half an hour.

\section{Background}
\label{sec:background}

This section introduces both compared benchmarking approaches in detail and highlights the specific characteristics when experimenting in cloud environments.

\subsection{Application Benchmarks}
\label{subsec:app_bench}

An application benchmark sets up the entire system under test (SUT) and takes a black-box perspective on it by stressing the SUT via its \emph{external} interfaces in a realistic way, i.e., similar to actual production traffic.
The benchmarking tool then observes the SUT's reactions on a range of Quality of Service (QoS) metrics, typically including performance~\cite{book_bermbach2017_cloud_service_benchmarking}.
In practice, application benchmarks create production-like workloads for either the complete SUT, e.g., to compare different database system alternatives, or specific parts of the SUT to reveal the performance of critical parts, e.g., studying the write scalability of a database system using a write-heavy workload.

Nowadays, software typically runs in cloud environments, which introduces new challenges for performance assessment~\cite{book_bermbach2017_cloud_service_benchmarking}.
Cloud VMs often have different performance characteristics and are subject to random fluctuations, even comparing two VMs of the same instance type~\cite{leitner2016patterns}.
Since we are only interested in a relative comparison of two SUT options and do not need absolute values, we can (largely) remove the noise resulting from cloud performance variability using \emph{duet benchmarking}~\cite{bulej2019initial,bulej2020duet}.
This is achieved, by running two (or even more) different SUT options, in our case different versions, and their application benchmarks on the same cloud VM simultaneously, with 50\% of the resources assigned to each SUT option and benchmark.
This ensures that measurements from all SUT options are subject to the exact same performance variability, i.e., the impact equally affects both terms of the equation when calculating the performance delta.

\subsection{Microbenchmarks}
\label{subsec:microbench}

Microbenchmarks do not deploy an entire SUT but instead evaluate the performance of single functions, thus, taking an \emph{internal} perspective.
They typically call a function repeatedly using randomized inputs, while gathering performance metrics such as execution time.
This can potentially reveal performance regressions quickly, while also pinpointing the functions which are the root cause of performance regression, but neglects more complex code interactions or the integration of software parts into the overall system.
For instance, calls to external systems are typically mocked similar to unit tests instead of deploying the external system.

Due to their lightweight nature, it has been proposed to use microbenchmarks in CI/CD pipelines to detect performance regressions automatically after a number of code changes, e.g.,~\cite{laaber2018evaluation,grambow2021using,grambow2022using}.
As such pipelines are typically executed on cloud VMs, similar considerations as discussed above are necessary to handle cloud performance variability.
Here, the current best practice is to use \emph{Randomized Multiple Interleaved Trials} (RMIT) execution~\cite{abedi2015conducting,abedi2017conducting} together with bootstrap analysis~\cite{kalibera2020quantifying, hesterberg2015what}.
The key idea is to repeat frequently and shuffle executions for each microbenchmark, e.g., repeating microbenchmark executions on different cloud VMs, shuffling the microbenchmark execution order, and repeating the execution of the microbenchmark suite multiple times.
Overall, this reduces measurement bias and ensures most accurate findings, but significantly increases execution time.

\section{Study Design}
\label{sec:study_design}

\begin{figure}
    \centering
    \includegraphics[width=0.8\columnwidth]{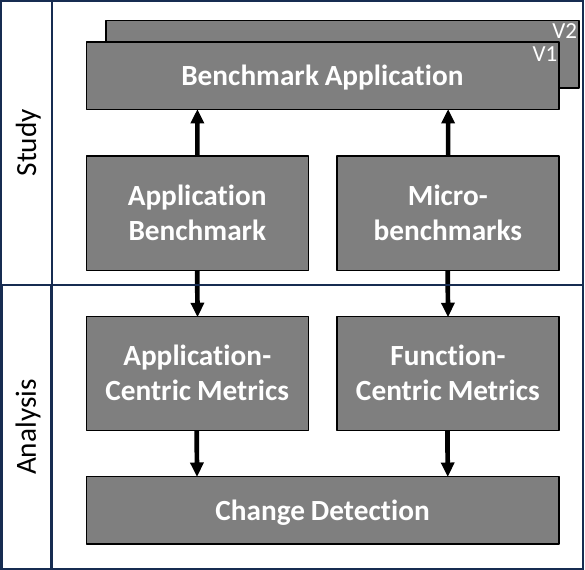}
    \caption{Overview of the study design.}
    \label{fig:architecture}
\end{figure}

In this section, we present the design of our study, which aims to evaluate the detection capabilities of application benchmarks and microbenchmark suites in identifying software performance issues.
For this, we first give an overview of our implemented testbed application, describe the three artificial performance issues, the designed application benchmark, the microbenchmark suite covering almost all source code parts, and finally outline our analysis approach.
\Cref{fig:architecture} gives a general high-level overview of the study design.

\subsection{Testbed Application}
Our testbed application is a small microservice application called \texttt{flight-booking-service}\footnote{\url{https://github.com/njapke/flight-booking-service}}.
We design this microservice-based application to provide the same functionality that a realistic booking service of an airline offers. 
Therefore, clients can look up flights, retrieve the available seats for a flight, and create new bookings.
The service performance can be studied by triggering such actions (application benchmark) or by executing single functions multiple times (microbenchmark).
The application has no external dependencies and stores its state in an in-memory key-value database that is based on \emph{BadgerDB}\footnote{\url{https://github.com/dgraph-io/badger}}. 
Here, the goal was to simplify the deployment and to minimize performance effects from external services. 

The testbed application exposes a RESTful API.
Using four endpoints, users can create flight booking (\path{/bookings}), search for destination airports (\path{/destinations}), search for flights (\path{/flights}), and reserve seats (\path{/flights/.../seats}).
In practice, these endpoints would be called from a web or mobile frontend, in our case they are called from the application benchmark.

The API implementation uses the \texttt{go-chi}\footnote{\url{https://go-chi.io/}} library as an HTTP router.
Furthermore, we use several middleware modules directly provided by the library (e.g., HTTP basic authentication or on-the-fly compression).

\subsection{Performance Issues}
\label{subsec:issues}
A key feature of our testbed application are the three performance issues with configurable severity.
As we wanted to allow others to easily replace the performance issues or to reuse them with a different testbed application, we implemented all three performance issues inside the \texttt{go-chi} library.
This way we measure the library's performance indirectly by using our flight booking microservice as a proxy.
To do so, we use a customized fork of the \texttt{go-chi} module in our microservice.
Each of our three implemented issues is on its own code branch (so the \texttt{main} branch always contains the unmodified version of our application) and the severity of the performance issues is configurable by setting an environment variable.
This enables the simulation of different performance scenarios and a precise evaluation of the system's response to varying levels of performance degradation.
In all cases, the severity ($s$) directly translates to the number of iterations a compute-intensive operation is repeated.
In order to guarantee a true A/A test at severity 0, we also include branches for all performance issues with a fixed severity 0 to compare against.

For studying the detection capabilities of both benchmark types, we include the following three artificial performance issues in the testbed application:

\paragraph{\textbf{A} -- Basic Auth Credential Validation:}
The basic auth middleware is part of the \texttt{go-chi} library, and it secures HTTP endpoints by requiring a valid username and password~\cite{Franks2019-basic-auth}.
We rely on the middleware to protect the \path{/bookings} endpoint and correctly link a newly created booking to a registered user account.
To validate the provided credentials, the middleware uses the \path{subtle.ConstantTimeCompare}\footnote{\url{https://pkg.go.dev/crypto/subtle\#ConstantTimeCompare}} function.
We implement this performance issue by replacing the original functionality with a less optimized solution.
The implementation hashes the input $s$ times using the SHA-512 hash function and calculates the deviation between the provided and the required password.
Thus, each validation requires more CPU resources and affects the application's overall performance more.

\paragraph{\textbf{B} -- Clean Path (Path Normalization):}
As part of the \texttt{go-chi} library, the clean path functionality helps to clean double slashes or double dots from the requested path, e.g., \path{/flights///myflightid} is cleaned to \path{/flights/myflightid}.
This allows the HTTP router to normalize typos and still be able to handle the request.
Internally, the middleware uses the Go standard library module \texttt{path} that provides a \texttt{Clean} function\footnote{\url{https://pkg.go.dev/path\#Clean}}.
In this performance issue, the \texttt{path.Clean} function is invoked $s$ times instead of once.
Thus, for every request that gets processed by the middleware, \texttt{path.Clean} is called multiple times and creates extra load for the CPU.
The expected performance impact is considered low, as the \texttt{path.Clean} function is highly optimized, especially if the path needs no normalization.
Furthermore, we use this functionality only for both \path{/flights/*} endpoints.
The performance issue will not affect the other endpoints (\path{/bookings} and \path{/destinations}).

\paragraph{\textbf{C} -- Request ID Generation:}
As part of the \texttt{go-chi} library, request ID generation assigns each incoming request a unique identifier.
This widespread practice links log messages to a specific request and makes debugging issues related to a particular request more manageable.
There, we implement our third performance issue:
We generate a request ID for every incoming request by reading $512*s$ bytes from the operating system's pseudo-random number generator and use the data to produce a message digest with the SHA-1 hash function.
This message digest is encoded as a hexadecimal string and populated as the request ID of the router's context.
The original implementation is much faster and uses a simple atomic counter to generate unique request IDs.
Hashing random data for every request will heavily impact the application's performance.

\subsection{Application Benchmark}
\label{subsec:appbench}
Following the duet benchmarking methodology, the application benchmark uses two cloud VMs, one for the SUT, the other for the workload generator. 
The first one, the \emph{application instance}, runs both versions of the application simultaneously.
Each version is bound to a fixed TCP port that serves the HTTP API.
Hence, the application instance exposes two ports which allows the benchmark client to access the running applications.
Furthermore, the available CPU capacity is limited for each version to 1.5 cores using the control groups (\emph{cgroups}) mechanism of the Linux kernel\footnote{\url{https://www.kernel.org/doc/html/latest/admin-guide/cgroup-v1/cgroups.html}}.

The second VM is the \emph{benchmark instance}, which is responsible for generating and sending the workload to both versions using benchmark clients.
There, we run two application benchmarks in parallel, one for each version running on the respective defined port.
For the implementation, we use Grafana \texttt{k6}\footnote{\url{https://k6.io/}}, which offers a closed workload model~\cite{book_bermbach2017_cloud_service_benchmarking}.
To cover the typical and realistic use cases of the application use case, we implemented two scenarios in the \texttt{k6} script:

\paragraph{\textbf{S}$_{\bm{1}}$ -- Flight Search:} The first scenario simulates a simple flight search and starts with retrieving a list containing all destination airports (\texttt{GET} \path{/destinations}). Next, we select a random airport and list all available flights that depart from this airport (\texttt{GET} \path{/flights?from=${airport}}).
\paragraph{\textbf{S}$_{\bm{2}}$ -- Flight Search \& Booking:} The second scenario extends the first one by combining a flight search with creating a new booking. For this, we pick a random flight and fetch the list of available seats for this flight (\texttt{GET} \path{/flights/${id}/seats}). Finally, we select two seats randomly and send a booking request containing the flight ID and seats to the application (\texttt{POST} \path{/bookings}). 

\bigskip

As the flight search is more common than creating a booking, we use different weights for both scenarios:
We configured \texttt{k6} to have a total of $50$ virtual users (VUs) with $2,000$ iterations per user for the flight search scenario and only $10$ VUs with $380$ iterations per user for the booking scenario.
This way, we run 103,800 flight searches (S$_1$, S$_2$) and additional 3,800 bookings (S$_2$) per benchmark run, thus, reflecting the application's real use case.
Initial trial experiments have shown that this setup using $60$ VUs in total utilizes all available CPU resources for both application versions on the \emph{application instance}.
A single execution of the application benchmark is about $30$ minutes long.
Since we will only focus on non-functional properties, we can ignore single request timeouts or HTTP error codes, even if they slightly influence the measurements.
Nevertheless, the overall measured performance change should remain the same as long as the amount of errors is balanced between both versions.

\begin{table}[t]
\centering
\begin{threeparttable}
\begin{tabularx}{\columnwidth}{clYYY}
\toprule
\multirow{2}{*}{\textbf{№}} & \multicolumn{1}{c}{\multirow{2}{*}{\textbf{Application Endpoint}}} & \multicolumn{3}{c}{\textbf{Perf.\ Issue}} \\ \cmidrule{3-5}
& \multicolumn{1}{c}{} & \phantom{\tnote{1}}A\tnote{1} & \phantom{\tnote{2}}B\tnote{2} & \phantom{\tnote{3}}C\tnote{3} \\ \midrule
E$_1$ & \makecell[l]{\texttt{POST /bookings}} & \ding{51} & \ding{55} & \ding{51} \\
E$_2$ & \makecell[l]{\texttt{GET\phantom{T} /destinations}} & \ding{55} & \ding{55} & \ding{51} \\
E$_3$ & \makecell[l]{\texttt{GET\phantom{T} /flights?from=\$\{airport\}}} & \ding{55} & \ding{51} & \ding{51} \\
E$_4$ & \makecell[l]{\texttt{GET\phantom{T} /flights/\$\{id\}/seats}} & \ding{55} & \ding{51} & \ding{51} \\ \bottomrule
\end{tabularx}
\begin{tablenotes}
    \item[1] Basic Auth
    \item[2] Clean Path
    \item[3] Request ID
\end{tablenotes}
\bigskip
\end{threeparttable}
\caption{Detection Capabilities of the Application Benchmark: Each of the three performance issues is detectable in a subset of the provided endpoints. We further label the endpoints with short names in the first column.}
\label{tab:endpoints-app-bench}
\end{table}

\Cref{tab:endpoints-app-bench} visualizes, which performance issues affect the four endpoints.
This also shows where the application benchmark could possibly detect the performance issues, as some endpoints should not show any different behavior for certain issues.

\subsection{Microbenchmark Suite}
The full microbenchmark suite contains $21$ individual microbenchmarks that cover all parts of our benchmark application.
We execute the suite to check whether any of the microbenchmarks detect the implemented performance issues.
For that purpose, we use the \emph{Randomized Multiple Interleaved Trials} (RMIT) methodology.
This allows us to fairly compare two versions of a microbenchmark by normalizing performance interference resulting from the underlying cloud infrastructure.
Hence, we run the microbenchmark suite three times and randomly change the benchmark execution order for every run, either execute the unaffected microbenchmark version first or the one with the performance issue, 
Also, each microbenchmark has two versions, we randomize the order in which we execute them and execute each microbenchmark five times for one second each\footnote{Microbenchmarks usually do not have fixed numbers of repetitions but are usually repeated for a fixed period of time.}.
Furthermore, we repeat this procedure on three different cloud instances resulting in $90$ measurements per microbenchmark, each including a one second period of repeatedly calling the function under test.

\begin{table}[t]
\centering
\setlength{\tabcolsep}{4pt} 
\begin{threeparttable}
\begin{tabularx}{\columnwidth}{cclYYY}
\toprule
\multirow{3}{*}{\textbf{G}} & \multirow{3}{*}{\textbf{№}} & \multicolumn{1}{c}{} & \multicolumn{3}{c}{\textbf{Perf.}} \\
& & \multicolumn{1}{c}{\textbf{Microbenchmark Name}} & \multicolumn{3}{c}{\textbf{Issue}} \\ \cmidrule{4-6}
& & \multicolumn{1}{c}{} & \phantom{\tnote{1}}A\tnote{1} & \phantom{\tnote{2}}B\tnote{2} & \phantom{\tnote{3}}C\tnote{3} \\ \midrule
1 & & database.Benchmark* & \ding{55} & \ding{55} & \ding{55} \\ \midrule
2 & & BenchmarkHandler* & \ding{55} & \ding{55} & \ding{55} \\ \midrule
& M$_1$ & \makecell[l]{BenchmarkRequestBookings} & \ding{51} & \ding{55} & \ding{51} \\
& M$_2$ & \makecell[l]{BenchmarkRequestCreateBooking} & \ding{51} & \ding{55} & \ding{51} \\
& M$_3$ & \makecell[l]{BenchmarkRequestDestinations} & \ding{55} & \ding{55} & \ding{51} \\
3 & M$_4$ & \makecell[l]{BenchmarkRequestFlight} & \ding{55} & \ding{51} & \ding{51} \\
& M$_5$ & \makecell[l]{BenchmarkRequestFlights} & \ding{55} & \ding{51} & \ding{51} \\
& M$_6$ & \makecell[l]{BenchmarkRequestFlightsQuery} & \ding{55} & \ding{51} & \ding{51} \\
& M$_7$ & \makecell[l]{BenchmarkRequestSeats} & \ding{55} & \ding{51} & \ding{51} \\ \bottomrule
\end{tabularx}
\begin{tablenotes}
    \item[1] Basic Auth
    \item[2] Clean Path
    \item[3] Request ID
\end{tablenotes}
\bigskip
\end{threeparttable}
\caption{Detection Capabilities of the Full Microbenchmark Suite: Each of the three performance issues is detectable by a set of microbenchmarks that cover the respective \texttt{go-chi} middleware. The first column shows different group membership of microbenchmarks. We further label the microbenchmarks of group 3 with short names in the second column, as only those have detection capabilities for the three performance issues.}
\label{tab:full-mb-suite}
\setlength{\tabcolsep}{6pt} 
\end{table}

We categorize the 21 microbenchmarks into three groups for a better understanding of what performance issues they should detect (see \Cref{tab:full-mb-suite}).
They contain seven microbenchmarks each and rank from the lowest coverage level to the highest, where the lowest means the respective benchmark covers only a small part of the application's codebase.
The first group of microbenchmarks focuses solely on the performance of the in-memory database implementation when inserting, reading, or searching objects (e.g., flights, bookings, etc.).
The second group measures the performance of the internal HTTP handlers.
This means they only benchmark the application's business logic without the \texttt{go-chi} HTTP router and its middleware.
The third and last group measures the performance through sequentially creating requests by calling the \texttt{ServeHTTP} function of the \texttt{go-chi} HTTP router.
This results in processing the HTTP requests fully and calling all involved middleware steps for the requested route.
All microbenchmarks create a deliberate overlap between different code paths to simulate realistic suites with redundancies~\cite{grambow2021using}. Thus, the injected performance issues can be detected by multiple microbenchmarks.
At first sight, this seems very similar to the application benchmark, as the microbenchmark covers the same code path while handling a similar request from the application benchmark.
However, this is incorrect because the application benchmark covers many more \emph{hidden} factors, e.g., network overhead and parallelism.
For example, the application performance might differ if multiple users try to create a flight booking simultaneously, compared to creating many bookings sequentially.

\subsection{Analysis}
To compare the application benchmark and microbenchmark suite, we first gather data for both types of benchmarks with different severity levels of each of the three performance issues.
Each experiment compares a base version with no performance issue ($v_1$) to a version with one of the three performance issues at a fixed severity level ($v_2$).
For the application benchmark, we use the overall request duration of requests to the four different REST endpoints as the performance metric, while for the microbenchmarks, we use the execution time of the particular microbenchmark.
In the remainder of this section, we use the term \emph{target} interchangeably for both the REST endpoints in the context of the application benchmark and the function called by a particular microbenchmark in the context of microbenchmarks -- we do this for reasons of readability.

Before the data is analyzed further, we need to take special precautions with the data from the application benchmark.
At the start of the experiment, the load that the application benchmark puts onto the SUT is still increasing.
As such, the SUT is not under full load at the start of the experiment.
This phase is called the warmup phase, as the SUT exhibits different performance characteristics than under full load, so data from this phase needs to be removed from the dataset used for analysis.
In our experiments, we remove the first $60$ seconds from the result data.
Due to using the duet benchmarking technique, a similar problem happens at the end of the experiment.
Since $v_1$ and $v_2$ are on the same VM, they only have comparable performance characteristics, when both versions are under full load.
One of the two versions might finish before the other, where the distance between the end of the application benchmark between both versions will only increase, if there are true performance changes between them.
Because of this, we also discard the last $60$ seconds before the first version finishes from the result data.

To detect performance changes, we calculate the \emph{median ratio} $r = \tilde{t}_{v2} / \tilde{t}_{v1}$ for each target, where $\tilde{t}_{v1}$, $\tilde{t}_{v2}$ are the median execution duration of a particular target in $v_1$ and $v_2$, respectively.
If $r = 1$, both versions take the same amount of median time to execute this particular target.
Should $r > 1$, then $v_2$ is slower than $v_1$, e.g., if $r = 1.3$ then $v_2$ takes 30\% longer on average than $v_1$ for this particular target.
Additionally, $r < 1$ shows $v_1$ being slower than $v_2$, with a similar interpretation as above.
Next, we calculate a \emph{confidence interval} (CI) for $r$ using the non-parametric method called bootstrap~\cite{kalibera2020quantifying, hesterberg2015what} with $10,000$ iterations and a confidence level of $99\%$.
Should this CI overlap 1, then the result cannot be statistically distinguished from $r = 1$, and therefore no performance change is detected.
Whenever this CI does not overlap 1, then we can determine that a performance change is present.
Should $v_2$ execute slower than $v_1$, we call the change a \emph{performance regression}.
Conversely, should $v_2$ execute faster than $v_1$, we call the change a \emph{performance improvement}.
A similar technique has been used by Laaber et al.\ for microbenchmarks in the past~\cite{laaber2021applying}.

Furthermore, we distinguish between \emph{small} and \emph{relevant} performance changes, where a small performance change has an impact less than or equal to $3\%$ ($r \in [0.97,1.03]$), and a relevant performance change has an impact greater than $3\%$ ($r > 1.03$ or $r < 0.97$).
This distinction is motivated by previous findings, as according to Georges~et~al.\ performance measurements often vary by about $3\%$~\cite{georges2007statistically} and Huang~et~al.\ consider performance regressions between $3\%$ and $20\%$ as relevant~\cite{mytkowicz2009producing}.
As such, it is often not possible to correctly identify small performance changes, which is why they are prone to being false positives.
For this reason, we differentiate both types of performance changes, as this affects the amount of false positives measured.

Finally, to give a few examples of this procedure: If we measure $r = 1.04$ with a CI not overlapping 1, we have a relevant performance regression of $4\%$.
Should we measure $r = 0.81$ with a wide CI overlapping 1, then we do not report a performance change, even though there could be a performance increase that is difficult to detect.

\section{Detecting Performance Issues}
\label{sec:results}

Using the testbed we proposed in \Cref{sec:study_design}, we now evaluate whether application benchmarks or microbenchmarks detect performance issues earlier.
First, we describe the general experiment setup and then report the benchmarking results for all three performance issues.
Finally, we outline our findings and the implications of our study.

\subsection{Experiment Setup}

For our experiments, we ran both micro- and application benchmarks on VMs in Google Cloud\footnote{\url{https://cloud.google.com}}.
The application benchmarks were executed on \texttt{n2-highcpu-4} instances, which have 4 vCPUs and 4 GB memory, for both the SUT and the load generator.
The microbenchmarks were executed on \texttt{n2-standard-2} instances with 2 vCPUs and 4 GB memory.

We ran $52$ experiments per performance issue, and tested $13$ severity levels for both the application benchmark and the microbenchmark suite, which we executed on three VMs for each severity level and three times in a row with RMIT execution mode.
In order to study a large range of severity levels, but also retain a high sampling rate at low severities, we ran our experiments at severities of powers of 2 plus A/A benchmarks using severity $0$.
Thus, our results have high resolution at low severity levels, and, conversely, low resolution at high severity levels.
We expect this to be a good trade-off decision, as a strong performance issue might reveal itself after only a few increases in severity in the lower levels, while a weak performance issue might reveal itself only in high severity levels, where only the order of magnitude of the severity contributes to measurable differences.
The highest severity level used is $2048$.

We used our \texttt{cloud-benchmark-conductor}\footnote{\url{https://github.com/njapke/cloud-benchmark-conductor}} to run all experiments.
It includes the application benchmark and automates all experiments, including running microbenchmark suites using RMIT.
Experiment configurations and results can be found in the same repository.

\subsection{Basic Auth}

\begin{table}[t]
    \centering
    \begin{threeparttable}
    \begin{tabularx}{\columnwidth}{>{\centering}p{0.6cm}YYYYYYYp{0pt}YYYY}
    \toprule
    & \multicolumn{7}{c}{Microbenchmarks Group 3} & & \multicolumn{4}{c}{\makecell{Application\\Endpoints}} \\\cmidrule(l{2pt}r{2pt}){2-8} \cmidrule(l{2pt}r{2pt}){10-13}
    Sev & \textbf{M}$_{\bm{1}}$ & \textbf{M}$_{\bm{2}}$ & M$_3$ & M$_4$ & M$_5$ & M$_6$ & M$_7$ & & \textbf{E}$_{\bm{1}}$ & E$_2$ & E$_3$ & E$_4$ \\
    \midrule
    0    & \emptybox & \emptybox & \emptybox & \emptybox & \emptybox & \emptybox & \emptybox &  & \emptybox & \emptybox & \emptybox & \emptybox \\
    1    & \emptybox & \emptybox & \emptybox & \emptybox & \emptybox & \emptybox & \emptybox &  & \emptybox & \emptybox & \emptybox & \emptybox \\
    2    & \emptybox & \hatchbox & \emptybox & \emptybox & \emptybox & \emptybox & \emptybox &  & \emptybox & \emptybox & \linesbox & \emptybox \\
    4    & \emptybox & \hatchbox & \emptybox & \emptybox & \emptybox & \emptybox & \emptybox &  & \emptybox & \emptybox & \sdotsbox & \emptybox \\
    8    & \emptybox & \hatchbox & \emptybox & \emptybox & \emptybox & \emptybox & \emptybox &  & \emptybox & \linesbox & \emptybox & \hatchbox \\
    16   & \hatchbox & \hatchbox & \emptybox & \emptybox & \emptybox & \emptybox & \emptybox &  & \emptybox & \emptybox & \emptybox & \emptybox \\
    32   & \hatchbox & \hatchbox & \emptybox & \emptybox & \emptybox & \emptybox & \emptybox &  & \emptybox & \emptybox & \emptybox & \emptybox \\
    64   & \hatchbox & \hatchbox & \emptybox & \emptybox & \emptybox & \emptybox & \emptybox &  & \emptybox & \emptybox & \sdotsbox & \emptybox \\
    128  & \hatchbox & \hatchbox & \emptybox & \emptybox & \emptybox & \emptybox & \emptybox &  & \hatchbox & \hatchbox & \hatchbox & \emptybox \\
    256  & \hatchbox & \hatchbox & \emptybox & \emptybox & \emptybox & \emptybox & \emptybox &  & \emptybox & \emptybox & \emptybox & \emptybox \\
    512  & \hatchbox & \hatchbox & \emptybox & \emptybox & \emptybox & \emptybox & \emptybox &  & \hatchbox & \emptybox & \linesbox & \emptybox \\
    1024 & \hatchbox & \hatchbox & \emptybox & \emptybox & \emptybox & \emptybox & \emptybox &  & \hatchbox & \emptybox & \emptybox & \emptybox \\
    2048 & \hatchbox & \hatchbox & \emptybox & \emptybox & \emptybox & \emptybox & \emptybox &  & \hatchbox & \hatchbox & \hatchbox & \emptybox \\
    \bottomrule
    \end{tabularx}
    \begin{tablenotes}
        \item \emptybox\hspace{4pt} no performance change
        \item \linesbox\hspace{4pt} $\leq3\%$ performance regression (small)
        \item \hatchbox\hspace{4pt} $>3\%$ performance regression (relevant)
        \item \sdotsbox\hspace{4pt} $\leq3\%$ performance increase (small)
        \item \ldotsbox\hspace{4pt} $>3\%$ performance increase (relevant)
    \end{tablenotes}
    \bigskip
    \end{threeparttable}
    \caption{Results for the Basic Auth performance issue.
    IDs of microbenchmarks/endpoints that could in principle detect the performance issue are marked in bold.
    M$_{\bm{1}}$ and M$_{\bm{2}}$ detect the performance issue at low severity levels while the application benchmark reports false positives and later detects the issue at E1 with severity level 512.}
    \label{tab:results-basic-auth}
\end{table}

\Cref{tab:results-basic-auth} shows the results for the Basic Auth performance issue.
M$_1$ and M$_2$ from group $3$ are the only microbenchmarks capable of detecting this performance issue and both detected the performance change.
M$_1$ reports the first performance change at severity 16, which is a performance regression of $5.1\%$ ($r=1.051$, CI: $[1.0279,1.1078]$), while M$_2$ reports the first performance change much earlier at severity 2 with a performance regression of $6.4\%$ ($r=1.064$, CI: $[1.0248,1.0917]$).
All other microbenchmarks gave expected results, except for one false positive in group 1, which showed a performance regression of 0.55\% (CI: $[1.0004,1.0133]$) at severity 1024.

The results of the application benchmark are a bit more mixed.
Only endpoint E$_1$ is capable of detecting the Basic Auth performance issue, but E$_2$, E$_3$, and E$_4$ also sporadically detect small and relevant performance regressions, while E$_3$ even detects small performance improvements at severities 4 and 64.
E$_1$ consistently identifies the performance issue only after severity 512, where the application benchmark reports a $20\%$ performance change ($r=1.2072$, CI: $[1.1014,1.3185]$).
The CI width for the application benchmark is generally larger than for the microbenchmarks, as it often exceeds $20\%$ (as in the aforementioned result for E$_1$), while for the microbenchmarks it typically does not exceed $10\%$.

\subsection{Clean Path}

\begin{table}[t]
    \centering
    \begin{threeparttable}
    \begin{tabularx}{\columnwidth}{>{\centering}p{0.6cm}YYYYYYYp{0pt}YYYY}
    \toprule
    & \multicolumn{7}{c}{Microbenchmarks Group 3} & & \multicolumn{4}{c}{\makecell{Application\\Endpoints}} \\\cmidrule(l{2pt}r{2pt}){2-8} \cmidrule(l{2pt}r{2pt}){10-13}
    Sev & M$_1$ & M$_2$ & M$_3$ & \textbf{M}$_{\bm{4}}$ & \textbf{M}$_{\bm{5}}$ & \textbf{M}$_{\bm{6}}$ & \textbf{M}$_{\bm{7}}$ & & E$_1$ & E$_2$ & \textbf{E}$_{\bm{3}}$ & \textbf{E}$_{\bm{4}}$ \\
    \midrule
    0    & \emptybox & \emptybox & \emptybox & \emptybox & \emptybox & \emptybox & \emptybox &  & \emptybox & \linesbox & \hatchbox & \emptybox \\
    1    & \emptybox & \emptybox & \emptybox & \hatchbox & \emptybox & \emptybox & \emptybox &  & \emptybox & \emptybox & \emptybox & \emptybox \\
    2    & \emptybox & \emptybox & \emptybox & \hatchbox & \emptybox & \emptybox & \emptybox &  & \emptybox & \emptybox & \sdotsbox & \emptybox \\
    4    & \emptybox & \emptybox & \emptybox & \hatchbox & \emptybox & \emptybox & \emptybox &  & \emptybox & \linesbox & \emptybox & \emptybox \\
    8    & \emptybox & \emptybox & \emptybox & \hatchbox & \emptybox & \emptybox & \emptybox &  & \emptybox & \emptybox & \linesbox & \emptybox \\
    16   & \emptybox & \emptybox & \emptybox & \hatchbox & \emptybox & \emptybox & \emptybox &  & \emptybox & \emptybox & \emptybox & \emptybox \\
    32   & \emptybox & \emptybox & \emptybox & \hatchbox & \emptybox & \emptybox & \emptybox &  & \emptybox & \ldotsbox & \ldotsbox & \emptybox \\
    64   & \emptybox & \emptybox & \emptybox & \hatchbox & \emptybox & \emptybox & \emptybox &  & \emptybox & \linesbox & \emptybox & \emptybox \\
    128  & \emptybox & \emptybox & \emptybox & \hatchbox & \emptybox & \emptybox & \emptybox &  & \emptybox & \emptybox & \emptybox & \emptybox \\
    256  & \emptybox & \emptybox & \emptybox & \hatchbox & \emptybox & \emptybox & \hatchbox &  & \emptybox & \emptybox & \emptybox & \emptybox \\
    512  & \emptybox & \emptybox & \emptybox & \hatchbox & \emptybox & \emptybox & \hatchbox &  & \emptybox & \hatchbox & \hatchbox & \emptybox \\
    1024 & \emptybox & \emptybox & \emptybox & \hatchbox & \linesbox & \emptybox & \hatchbox &  & \emptybox & \linesbox & \emptybox & \emptybox \\
    2048 & \emptybox & \emptybox & \emptybox & \hatchbox & \hatchbox & \emptybox & \hatchbox &  & \emptybox & \hatchbox & \hatchbox & \emptybox \\
    \bottomrule
    \end{tabularx}
    \begin{tablenotes}
        \item \emptybox\hspace{4pt} no performance change
        \item \linesbox\hspace{4pt} $\leq3\%$ performance regression (small)
        \item \hatchbox\hspace{4pt} $>3\%$ performance regression (relevant)
        \item \sdotsbox\hspace{4pt} $\leq3\%$ performance increase (small)
        \item \ldotsbox\hspace{4pt} $>3\%$ performance increase (relevant)
    \end{tablenotes}
    \bigskip
    \end{threeparttable}
    \caption{Results for the Clean Path performance issue.
    IDs of microbenchmarks/endpoints that could in principle detect the performance issue are marked in bold.
    M$_{\bm{4}}$ already detects this performance issue at severity level 1.
    The application benchmark can not detect this low-impact issue reliably.}
    \label{tab:results-clean-path}
\end{table}

\Cref{tab:results-clean-path} shows the results for the Clean Path performance issue.
M$_4$, M$_5$, M$_6$ and M$_7$ are the four microbenchmarks capable of detecting this performance issue, although M$_6$ does not report a performance change for any of the benchmarked severity levels.
This is likely due to the overall low impact of this performance issue, as mentioned in \Cref{subsec:issues}.
M$_4$, however, already reports a relevant performance regression at severity 1 of $3.85\%$ ($r=1.0385$, CI: $[1.0228,1.054]$).
M$_5$ and M$_7$ reported performance regressions only at higher severities, with M$_7$ being earlier at severity $256$ with $3.08\%$ ($r=1.0308$, CI: $[1.0155,1.061]$), and M$_5$ being late at severity $2048$ with the first relevant performance change of $6.66\%$ ($r=1.0666$, CI: $[1.0416,1.0761]$).
There were no false positives, i.e., no unaffected microbenchmarks that reported a performance change.

The results of the application benchmark, however, are all over the place.
Endpoints E$_3$ and E$_4$ should be able to detect the Clean Path performance issue, but neither detects it reliably.
E$_4$ does not report any changes, while E$_3$ sporadically detects small and relevant performance improvements and regressions.
Only E$_1$ correctly reports no performance changes.
Even E$_2$ sporadically detects performance improvements and regressions, even though it should report no changes.
Most notably, the A/A test at severity 0 fails for both E$_2$ and E$_3$, where both detect a performance regression.

\subsection{Request ID}

\begin{table}[t]
    \centering
    \begin{threeparttable}
    \begin{tabularx}{\columnwidth}{>{\centering}p{0.6cm}YYYYYYYp{0pt}YYYY}
    \toprule
    & \multicolumn{7}{c}{Microbenchmarks Group 3} & & \multicolumn{4}{c}{\makecell{Application\\Endpoints}} \\\cmidrule(l{2pt}r{2pt}){2-8} \cmidrule(l{2pt}r{2pt}){10-13}
    Sev & \textbf{M}$_{\bm{1}}$ & \textbf{M}$_{\bm{2}}$ & \textbf{M}$_{\bm{3}}$ & \textbf{M}$_{\bm{4}}$ & \textbf{M}$_{\bm{5}}$ & \textbf{M}$_{\bm{6}}$ & \textbf{M}$_{\bm{7}}$ & & \textbf{E}$_{\bm{1}}$ & \textbf{E}$_{\bm{2}}$ & \textbf{E}$_{\bm{3}}$ & \textbf{E}$_{\bm{4}}$ \\
    \midrule
    0    & \emptybox & \emptybox & \emptybox & \linesbox & \emptybox & \emptybox & \emptybox &  & \emptybox & \emptybox & \emptybox & \emptybox \\
    1    & \emptybox & \hatchbox & \emptybox & \hatchbox & \emptybox & \emptybox & \emptybox &  & \emptybox & \emptybox & \linesbox & \emptybox \\
    2    & \emptybox & \hatchbox & \emptybox & \hatchbox & \emptybox & \emptybox & \emptybox &  & \emptybox & \emptybox & \emptybox & \emptybox \\
    4    & \emptybox & \hatchbox & \emptybox & \hatchbox & \emptybox & \emptybox & \linesbox &  & \emptybox & \linesbox & \linesbox & \emptybox \\
    8    & \hatchbox & \hatchbox & \emptybox & \hatchbox & \emptybox & \emptybox & \hatchbox &  & \emptybox & \linesbox & \linesbox & \emptybox \\
    16   & \hatchbox & \hatchbox & \emptybox & \hatchbox & \hatchbox & \linesbox & \hatchbox &  & \emptybox & \hatchbox & \hatchbox & \emptybox \\
    32   & \hatchbox & \hatchbox & \emptybox & \hatchbox & \hatchbox & \emptybox & \hatchbox &  & \hatchbox & \hatchbox & \hatchbox & \hatchbox \\
    64   & \hatchbox & \hatchbox & \linesbox & \hatchbox & \hatchbox & \linesbox & \hatchbox &  & \hatchbox & \hatchbox & \hatchbox & \hatchbox \\
    128  & \hatchbox & \hatchbox & \hatchbox & \hatchbox & \hatchbox & \hatchbox & \hatchbox &  & \hatchbox & \hatchbox & \hatchbox & \hatchbox \\
    256  & \hatchbox & \hatchbox & \hatchbox & \hatchbox & \hatchbox & \hatchbox & \hatchbox &  & \hatchbox & \hatchbox & \hatchbox & \hatchbox \\
    512  & \hatchbox & \hatchbox & \hatchbox & \hatchbox & \hatchbox & \hatchbox & \hatchbox &  & \hatchbox & \hatchbox & \hatchbox & \hatchbox \\
    1024 & \hatchbox & \hatchbox & \hatchbox & \hatchbox & \hatchbox & \hatchbox & \hatchbox &  & \hatchbox & \hatchbox & \hatchbox & \hatchbox \\
    2048 & \hatchbox & \hatchbox & \hatchbox & \hatchbox & \hatchbox & \hatchbox & \hatchbox &  & \hatchbox & \hatchbox & \hatchbox & \hatchbox \\
    \bottomrule
    \end{tabularx}
    \begin{tablenotes}
        \item \emptybox\hspace{4pt} no performance change
        \item \linesbox\hspace{4pt} $\leq3\%$ performance regression (small)
        \item \hatchbox\hspace{4pt} $>3\%$ performance regression (relevant)
        \item \sdotsbox\hspace{4pt} $\leq3\%$ performance increase (small)
        \item \ldotsbox\hspace{4pt} $>3\%$ performance increase (relevant)
    \end{tablenotes}
    \bigskip
    \end{threeparttable}
    \caption{Results for the Request ID performance issue.
    IDs of microbenchmarks/endpoints that could in principle detect the performance issue are marked in bold.
    All microbenchmarks detect this performance issue.
    The application benchmark is also able to detect this high-impact issue reliably.}
    \label{tab:results-request-id}
\end{table}

\Cref{tab:results-request-id} shows the results for the Request ID performance issue.
This performance issue can be detected by all microbenchmarks in group 3.
All microbenchmarks of group $3$ correctly detect the performance issue once a certain severity level is reached.
M$_2$ and M$_4$ detect the issue the earliest at severity 1.
M$_2$ detects a performance regression of $8.74\%$ ($r=1.0874$, CI: $[1.0557,1.1203]$), while M$_4$ detects a performance regression of $55.89\%$ ($r=1.5589$, CI: $[1.5338,1.5937]$) at this severity level.
Notably, M$_4$ also detects a small performance regression during the A/A test at severity 0.
As small performance changes are often false positives, this is not an unusual occurrence.
All microbenchmarks besides group 3 correctly reported no performance changes.

The application benchmark identifies this performance regression better than the other issues, as the Request ID performance issue has a much stronger overall impact.
All endpoints are capable of detecting this performance issue.
Endpoints E$_2$ and E$_3$ both continuously detect a relevant performance regression after severity 16, which is slightly earlier than E$_1$ and E$_4$ at severity 32.
The measured performance regressions at severity 16 are $4.58\%$ ($r=1.0458$, CI: $[1.0293,1.0631]$) for E$_2$, and $4.51\%$ ($r=1.0451$, CI: $[1.0291,1.0622]$) for E$_3$.
The measured performance regressions at severity 32 are $14.78\%$ ($r=1.1478$, CI: $[1.0371,1.2395]$) for E$_1$, and $16.28\%$ ($r=1.1628$, CI: $[1.0595,1.2578]$) for E$_4$.

\begin{figure}
    \centering
    \includegraphics[width=\columnwidth]{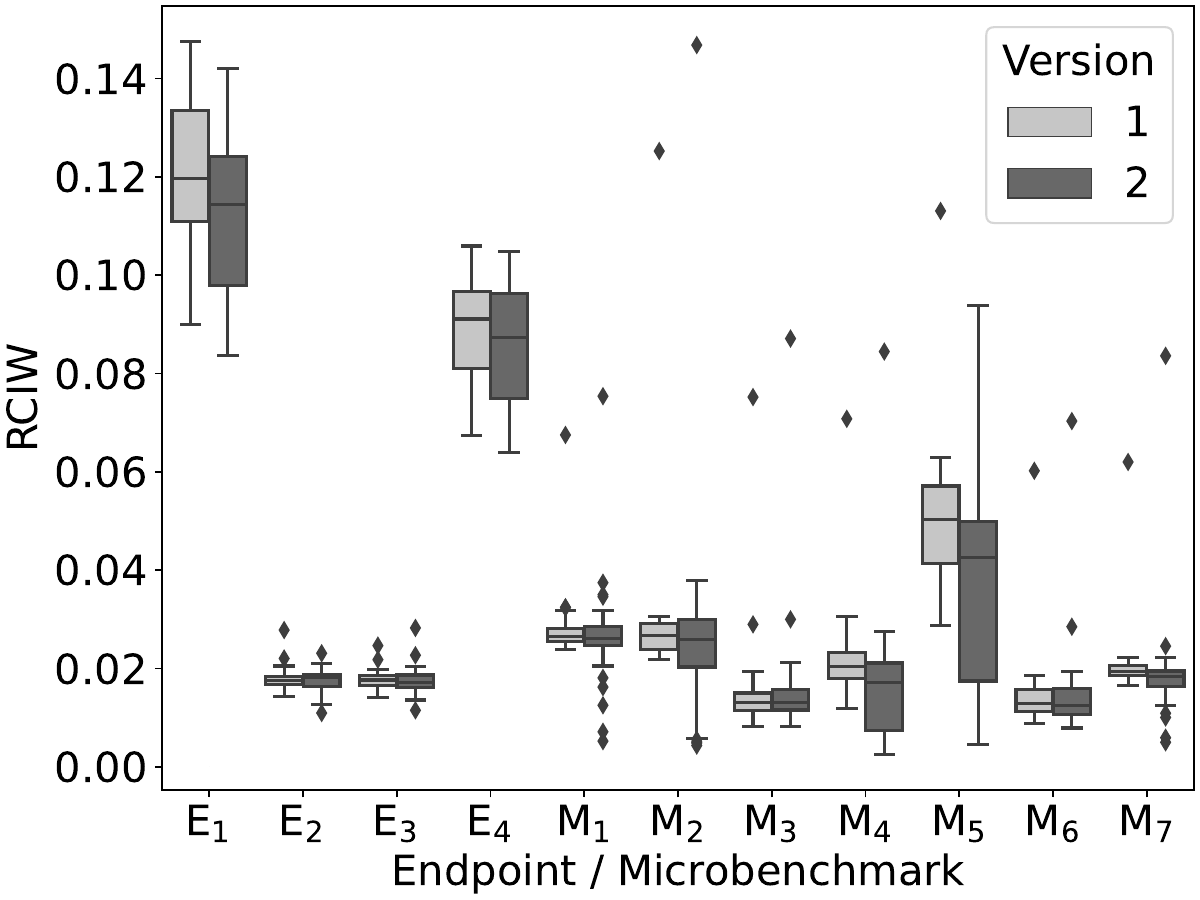}
    \caption{This figure visualizes the RCIW distribution across experiments for each endpoint and microbenchmark as boxplots.}
    \label{fig:rciw_boxplot}
\end{figure}

\subsection{Accuracy Analysis}
As alluded to before, the width of CIs for some application endpoints are larger, leading only to reported performance changes if the change has a sufficiently high magnitude.
\Cref{fig:rciw_boxplot} shows the \emph{relative confidence interval width} (RCIW) for all application benchmark endpoints and microbenchmarks of group 3.
The RCIW is the width of a CI, which is normalized by dividing by a statistical average.
Here, we calculated the CI for the median request latency for the application benchmark endpoints, and for the median execution duration for the microbenchmarks.
In both cases, we used bootstrap with $10,000$ iterations and a confidence level of $99$ and divided by the median to calculate the RCIW.
This way, each experiment run yields one RCIW for each endpoint/microbenchmark.
We repeated this procedure for every performance issue and severity level, giving us a distribution of the RCIW for each endpoint and microbenchmark, which is visualized by the boxplots.
They clearly show, that endpoints E$_1$ and E$_4$ tend to have the widest CIs, while endpoints E$_2$ and E$_3$ have the most narrow CIs and should be able to detect performance issues earlier.
Interestingly, the narrow CIs of E$_2$ and E$_3$ might also increase the likelihood of false positives.
These endpoints can pick up changes so well, that differences due to noise are picked up as statistically significant.
The microbenchmarks tend to have more narrow CIs, but with some higher outliers.

\subsection{Findings and Implications}
We now report on effectiveness and efficiency of using microbenchmarks or application benchmarks for detecting performance regressions.

\paragraph{Effectiveness}
The microbenchmarks were more effective at finding performance issues than the application benchmark, as some microbenchmarks that benchmarked functions directly affected by the performance issues could identify a performance regression well, even at low severities.
The application endpoints, which were stressed by the application benchmark, did not give conclusive results for low impact performance issues, and only identified high impact performance issues, such as the Request ID issue, but not as early as the microbenchmarks did.
Both results are to be expected since the relative latency impact of a performance regression tends to be higher for benchmarks which measure low latency values (typical microbenchmark scenarios) than for benchmarks measuring higher latency values (typical application benchmark scenarios which measure end-to-end latency).
Furthermore, two out of four endpoints tended to have wide CIs, meaning they can generally only identify higher impact performance issues.
This could be remedied by collecting more data, which generally narrows CIs.
As application benchmarks evaluate the SUT in a manner closer to production usage, they are better equipped to quantify the actual production impact of performance issues.
Microbenchmarks may find performance issues earlier, but they cannot quantify how that will impact the production use of the software, e.g., the microbenchmark suite might indicate that a security patch slows down a certain function, but without an application benchmark it is impossible to quantify the overall impact.
There are approaches for calculating the impact of a performance regression in a microbenchmark to the overall impact in an application benchmark, such as the \emph{reference impact} by Grambow et al.~\cite{grambow2022using}.

\paragraph{Efficiency}
While the microbenchmarks may have been more effective at uncovering performance issues, it can take very long to execute a large microbenchmark suite that covers the entire source code well.
In our experiments, executing the entire microbenchmark suite using an RMIT setup of 3 instance runs, 3 suite runs, 5 iterations, and running each iteration for 1 second took about one hour.
This includes VM startup, as well as setup and execution.
The application benchmark, however, needed only two VMs due to duet benchmarking, and finished within half an hour.
This means that, while a microbenchmark suite that covers the source code well is highly effective, it is not as efficient as an application benchmark.

\paragraph{Implications}
In practice, both approaches should be combined to receive early indications of performance regressions, and to accurately quantify their production impact.
Application benchmarks are better at detecting performance regressions which will matter in production, cannot pinpoint the source of the regressions (beyond a git diff between the versions compared), but are significantly faster and less expensive to execute.
Microbenchmarks can detect performance regressions earlier and also help to pinpoint their source (since we know the function targeted by the microbenchmark) but it will usually be unclear whether the regression matters for production, the benchmark execution will take longer, and is overall more expensive.

It should also be noted that our testbed application would usually be a single microservice in a larger application.
In terms of benchmarking, this means that a larger application would still use the same duration for the application benchmark (though probably running on more VMs in parallel~\cite{paper_grambow2020_benchmarking_microservices,paper_grambow_benchmarking_2020}) whereas the number of microbenchmarks will scale with the number of other microservices.

\section{Discussion}
\label{sec:discussion}
In this section, we discuss threats to validity and other limitations of our findings.

\paragraph{Threats to Validity}
The main threat to construct validity concerns the methods we used for quantifying performance changes.
We use a median ratio of either request latency (application benchmark), or execution duration (microbenchmarks) to quantify a performance change.
Furthermore, we use bootstrap confidence intervals to test, whether a performance change is statistically significant.
There are alternative metrics and techniques for quantifying performance changes which are likely to give different results.
Generally, the combination we used is in line with the state of the art and best practices in benchmarking:
Bootstrap confidence intervals are state of the art and have often been used by practitioners of performance engineering~\cite{kalibera2020quantifying,laaber2021applying,laaber2019software}, while the median ratio is inspired by the similar mean ratio used by Laaber et al.\ for quantifying performance changes~\cite{laaber2021applying}.
Furthermore, we preprocess the experiment data from the application benchmark by applying a median operator to each second of the experiment.
This way, the amount of data is reduced to one data point per second in each experiment.
The advantage of this procedure is faster calculation of bootstrap CIs.
Such preprocessing steps can always influence the final result, but in this case, we ran a preliminary analysis using the original data, and found the changes to be negligible.

Threats to internal validity mainly concern influences on performance measurements.
All our experiments were run on GCP Compute instances, which, as Cloud VMs, suffer from unreliable performance.
We largely mitigate this using duet benchmarking~\cite{bulej2019initial,bulej2020duet} and RMIT~\cite{abedi2017conducting,abedi2015conducting}, both of which are state of the art when working with Clouds~\cite{grambow2022using,laaber2019software}.
Some effects will, however, remain as can also be seen in the false positives we reported.

Threats to external validity concern our implementation of the benchmarking testbed, the flight booking service.
While ultimately all of our findings are only directly applicable to the testbed used and our implementation of benchmarks, we designed both according to be representative of real applications.
Generally, it would have been possible to use a real cloud application.
In practice, these are, however, either not accessible to the public, are lacking a fitting application benchmark, or are missing a microbenchmark suite with full code coverage.

\paragraph{Results are Limited to Particular Setup}
While we have taken care to ensure our results are as generally applicable as possible, ultimately they are limited to the particular setup of our study.
Most insights will only be applicable to application benchmarks using duet benchmarking and microbenchmarks using RMIT.
Furthermore, the design of our testbed, the flight-booking-service, as well as the use of the Go programming language, influences the results, making it difficult to apply the insights to other programming languages and environments.

\paragraph{Influence of Duet Benchmarking and RMIT}
The results clearly showed that the application benchmark often measured false positives, i.e., relevant performance changes that do not exist in the code.
These false positives are often measured for multiple endpoints during the same experiment.
In \Cref{tab:results-basic-auth,tab:results-clean-path,tab:results-request-id}, each line corresponds to one experiment for the application benchmark, and the microbenchmark suite, meaning all microbenchmarks share the same three VMs, and all endpoints share the same two VMs, if they appear in the same line.
It is immediately apparent that multiple false positives appear during the same experiment, leading to the conclusion that results are biased due to platform influence.
This influence can be removed by repeating the experiments with different VMs several times which, however, significantly increases cost.
As RMIT uses multiple repetitions across VMs, such influences are much smaller for the microbenchmark results, leading only to a single small performance change being measured that is a false positive.
Overall, this will likely result in a trade-off between effectiveness and efficiency: Additional repetitions increase effectiveness, leading to better results, but decrease efficiency, as each additional repetition costs time and money.

\paragraph{Benchmark Application not used in Production}
While we took care in designing the flight-booking-service as realistically as possible, it has never been run in actual production environments.
The process of using a microservice in production also determines the direction of development, especially regarding potential external dependencies.
For ease of use in experiments, the flight booking service does not have any external dependencies, and uses an in-memory database.
A real-world implementation would likely use an external database system or service, as well as many more microservices for more complex functionality.
Finally, we artificially inject performance regressions but do so in an as realistic as possible way.
While it may have been preferable to study real performance regressions, we would not have been able to study how different configurable severity levels affect how well the two benchmark types can detect performance issues.

\section{Related Work}
\label{sec:related_work}

Our testbed application uses three artificially injected performance issues with adjustable severity. 
Other studies analyzing performance regressions, however, investigate performance issues in production systems~\cite{zaman_qualitative_2012, chen_exploratory_2017}. 
Moreover, there are also several approaches to automatically identify the root cause of a performance issue~\cite{heger_automated_2013, nguyen_industrial_2014}.
Using a production system with real performance issues in our study was not possible since there is no project matching all our requirements.
Implementing our own performance testbed allowed us to study both benchmark types for different severity levels of each issue in detail.

Performance issues should ideally be detected after a code change is committed.
Similar to our study, there are several other studies motivating to use automated benchmarking in a cloud environment for this~\cite{grambow_continuous_2019, javed_perfci_2020}.
Even though several studies show that benchmarks in cloud environment are hard to implement due to variability of measurements and random fluctuations~\cite{leitner2016patterns,laaber2019software}, we could detect some regressions already at severity level 1 in our experiments. 

In our study, we always executed the complete microbenchmark suite without further optimizations and thus had a suite execution time of about one hour. 
The suite execution, however, can be optimized by dynamically stopping microbenchmarks when there are stable results~\cite{he_statistics-based_2019}, adjust the execution order of microbenchmarks within the suite~\cite{mostafa2017perfranker,laaber2021applying,laaber2022multi}, or selecting and execution only a subset of relevant microbenchmarks~\cite{de_oliveira_perphecy_2017,grambow2021using}.
Our testbed application with a microbenchmark suite covering almost all source code parts can also be used to evaluate these optimization approaches.

Furthermore, our performance issue testbed application can also be used to evaluate performance change point detection algorithms. 
While our study used a threshold-based detection, it is also possible to apply and study other algorithms~\cite{daly_industry_2019,fleming2023hunter}.

Finally, there are a number of application benchmarks specifically made for cloud environments, e.g.~\cite{paper_bermbach2017_benchfoundry,cooper2010benchmarking,borhani2014wpress,paper_difallah_oltpbench}.
We believe that our findings from this paper can be used to shed some light on the strengths and weaknesses of these approaches in the context of detecting performance regressions across software versions.

\section{Conclusion}
\label{sec:conclusion}

In this paper, we presented an open-source benchmarking testbed, which includes a flight booking microservice as the SUT, an application benchmark, and a microbenchmark suite, and which supports injection of three different performance issues with customizable severity.
We collected an extensive data set of benchmarking experiments using the application benchmark and the microbenchmark suite to study how well both benchmark types can detect performance regressions.
We showed that microbenchmarks typically detect the performance issues much earlier than the application benchmark.
The microbenchmarks are, however, not able to assess production-like behavior of the SUT, as they only cover small sections of code each, and the entire suite has a longer execution duration, leading to significantly lower efficiency of this approach.

\balance
\bibliographystyle{ACM-Reference-Format}
\bibliography{bibliography}

\end{document}